\documentclass[10pt, a4paper, twocolumn, twoside]{article} % 10pt font size (11 and 12 also possible), A4 paper (letterpaper for US letter) and two column layout (remove for one column)

\usepackage[english]{babel} % English language hyphenation

\usepackage{microtype} % Better typography

\usepackage{amsmath,amsfonts,amsthm} % Math packages for equations

\usepackage[svgnames]{xcolor} % Enabling colors by their 'svgnames'

\usepackage[small, labelfont=bf, up]{caption} % Custom captions under/above tables and figures

\usepackage{booktabs} % Horizontal rules in tables

\usepackage{lastpage} % Used to determine the number of pages in the document (for "Page X of Total")

\usepackage{graphicx} % Required for adding images

\usepackage{enumitem} % Required for customising lists
\setlist{noitemsep} % Remove spacing between bullet/numbered list elements

\usepackage{sectsty} % Enables custom section titles
\allsectionsfont{\usefont{OT1}{phv}{b}{n}} % Change the font of all section commands (Helvetica)

\usepackage{geometry} % Required for adjusting page dimensions

\usepackage[T1]{fontenc} % Output font encoding for international characters
\usepackage[utf8]{inputenc} % Required for inputting international characters

\usepackage{XCharter} % Use the XCharter font

\usepackage{hyperref}

\usepackage{journals}

\usepackage{fancyhdr} % Needed to define custom headers/footers
\newcommand{\shorttitle}[1]{\fancyhead[CE]{\textsl{#1}}}
\newcommand{\shortauthors}[1]{\fancyhead[CO]{\textsl{#1}}}

\fancypagestyle{firstpage}{ % Page style for the first page with the title
	\fancyhf{}
        \lhead{\vspace*{0.0em} \textit{\footnotesize 21$^{\rm st}$ European Workshop on White
            Dwarfs \\ Castanheira, Campos, Montgomery, eds. \\[-0.2em]
          July 23--27, 2018, Austin, Texas, USA}}
}

\date{}

\newcommand{\authorstyle}[1]{{\large\usefont{OT1}{phv}{b}{n}\color{DarkRed}#1}} % Authors style (Helvetica)

\newcommand{\institution}[1]{{\footnotesize\usefont{OT1}{phv}{m}{sl}\color{Black}#1}} % Institutions style (Helvetica)
\usepackage{titling} % Allows custom title configuration

\newcommand{\HorRule}{\color{DarkGoldenrod}\rule{\linewidth}{1pt}} % Defines the gold horizontal rule around the title

\pretitle{
	\vspace{-30pt} % Move the entire title section up
	\HorRule\vspace{10pt} % Horizontal rule before the title
	\fontsize{16}{12}\usefont{OT1}{phv}{b}{n}\selectfont % Helvetica
	\color{DarkRed} % Text colour for the title and author(s)
}

\posttitle{\par\vskip 10pt} % Whitespace under the title

\preauthor{} % Anything that will appear before \author is printed

\postauthor{ % Anything that will appear after \author is printed
	\vspace{0pt} % Space before the rule
	{\par\HorRule} % Horizontal rule after the title
	\vspace{-10pt} % Space after the title section
}

\newcommand{\newabstract}[1]{
    {\section*{Abstract}
    \bfseries #1}
  }

\usepackage[authoryear]{natbib}
\title{White dwarf collisions and the meteoritic Ne-E annomaly} % The article title

\shorttitle {White Dwarfs and Ne-E } % The short article title for page headings

\shortauthors{Isern, Bravo} % The short author list for page headings

\author{
        \authorstyle{J.~Isern,$^{1,2,3}$, and E.~Bravo$^4$}
	\newline\newline % Space before institutions
	$^1$\institution{Institut de Ci\`encies de l'Espai (ICE,CSIC), Campus UAB, 08123 Cerdanyola del Vall\`es, Spain; 
          isern@ice.cat}\\ % Institution 1
	$^2$\institution{Institut d'Estudis Espacials de Catalunya (IEEC), 08028  Barcelona, Spain}\\  
         $^3$\institution{Reial Acad\`emia de Ci\`encies i Arts de Barcelona (RACAB)}\\
	$^4$\institution{Dept. de F\' {\i}sica, ETS Arquitextura del Vall\`es (UPC), 08173 Sant Cugat del Vall\`es, Spain;  eduardo.bravo@upc.edu} % Institution 3
      }

\begin{document}

\maketitle % Print the title

\thispagestyle{firstpage} % Apply the page style for the first page

%----------------------------------------------------------------------------------------
%	ABSTRACT
%----------------------------------------------------------------------------------------

\newabstract{
The analysis of noble gases in primitive meteorites has shown the existence of anomalous isotopic abundances when compared with the average Solar System values. In particular it has been found that some graphite grains contain a unexpected high abundance of neon-22. This excess of neon-22 is usually attributed to the radioactive decay of sodium-22 produced in the O/Ne burning layer of a core collapse supernova. In this talk we speculate about a different origin, the disruption of a crystallized white dwarf by a compact object (white dwarf, neutron star or black hole).

  }

%----------------------------------------------------------------------------------------
%	ARTICLE BODY
%----------------------------------------------------------------------------------------

\section{Introduction}
For some time it was believed that white dwarf collisions were rare encounters that only could occur in dense ambients like the core of globular clusters or the regions surrounding the central black hole of galaxies, for which reason its importance as nucleosynthesis agents was disregarded \citep{benz89}. This perception changed when it was realized that the majority of stars are not isolated but forming part of multiple (binary, ternary ...) systems.

Multiple systems are in general unstable except those that have an hierarchical structure, being a typical example a binary system plus a third star. These systems, however, can exchange angular momentum and produce periodical modifications of the eccentricity of the binary and of the inclination of the inner and outer orbital planes (Kodai-Lidov cycles) leading to collisions or at least strong encounters that, in the case of two white dwarfs and depending on the circumstances, can induce thermonuclear explosions and ejections of pristine matter. In any case, the evolution of stars in such conditions is not well understood \citep{toon16}.

The frequency of such collisions is an open question, \citet{katz12} suggested that the number of frontal collisions could be as large that this mechanism alone could be responsible of a large fraction of Type Ia supernovae (SNIa). This idea has been challenged by \citet{toon18} who, through detailed triple star evolution population synthesis, found that isolated triples can only account for a small fraction of SNIa. In any case it is important to realise that the number of failed white dwarf-white dwarf collisions is larger than previously expected and, specially, that they are not confined to the core of globular cluster nor to the outskirts of the central black hole.

This means that if the ejected material, either with the primitive composition of the white dwarf or with a composition modified by the nucleosynthesis associated to the collision, can preserve its composition in the form of stardust and be incorporated to a pre-stellar nebula could be responsible of chemical anomalies similar to those found in the primitive Solar System meteorites \citep{iser18}.

\section{Chemical structure of the white dwarf}
White dwarfs are the remnants of low and intermediate-mass stars and, as a consequence of the pressure of degenerate electrons, they cannot synthesize new atomic nuclei. Therefore, when they start the cooling process, their composition is completely similar to that of the AGB cores.  Furthermore, during the cooling the isotopic distribution changes as a consequence of sedimentation and, depending on the strength, of the nucleosynthesis induced by the shock. Thus, the material injected to the interstellar medium (ISM) material has a chemical composition completely peculiar.

\begin{figure}[t]
  \centerline{\includegraphics[width=1.0\columnwidth]{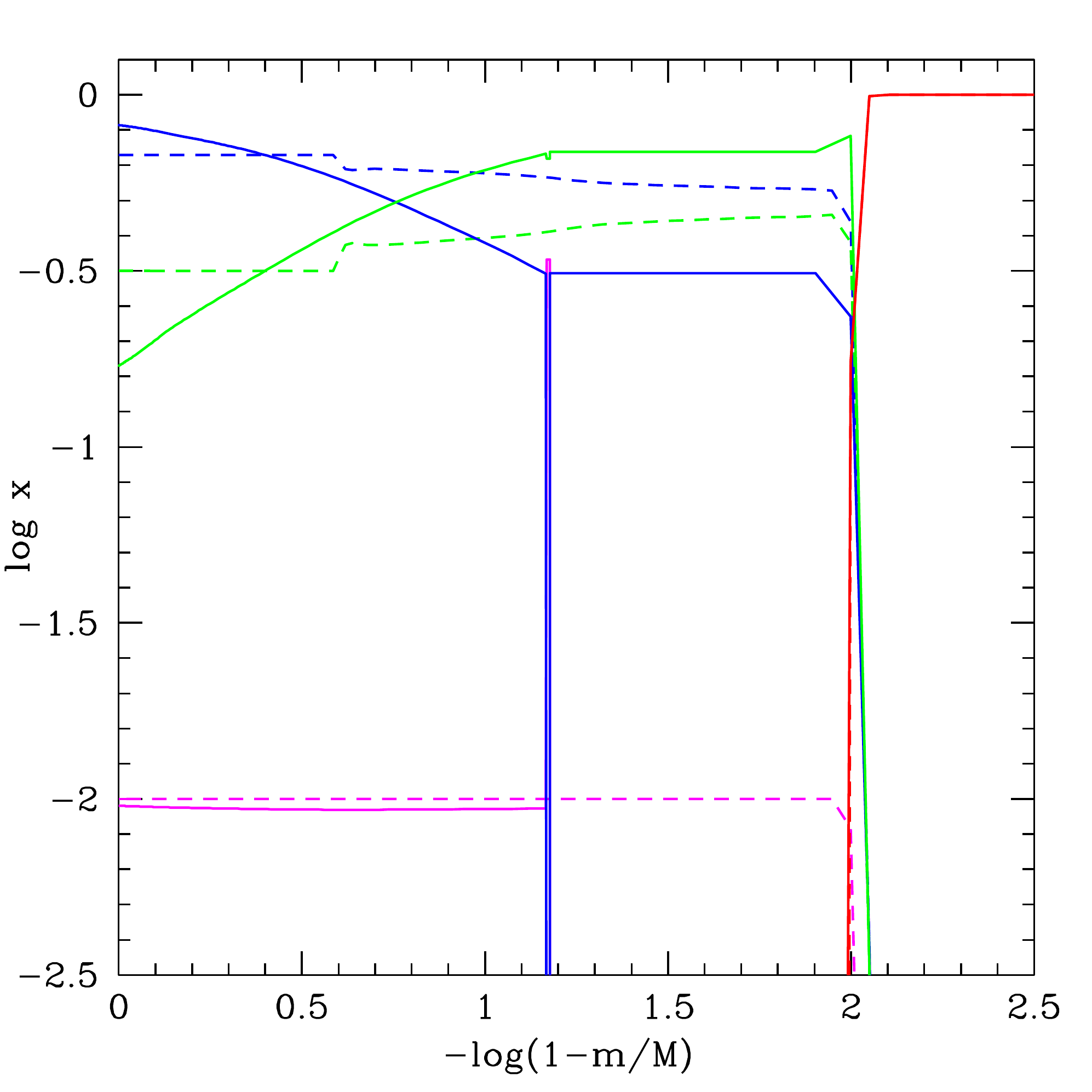}}
  \caption{Chemical profile of a typical C/O white dwarf as a function of the mass coordinate, where m and M are, respectively the mass coordinate and the total mass. Dashed lines display the initial configuration assuming it is made only of $^{12}$C (green), $^{16}$O (blue) and $^{22}$Ne (magenta). Solid lines represent the chemical composition that is obtained after complete crystallization using the phase diagram of \citet{segr96}.  } 
  \label{fig1}
\end{figure}

\begin{figure}[t]
  \centerline{\includegraphics[width=1.0\columnwidth]{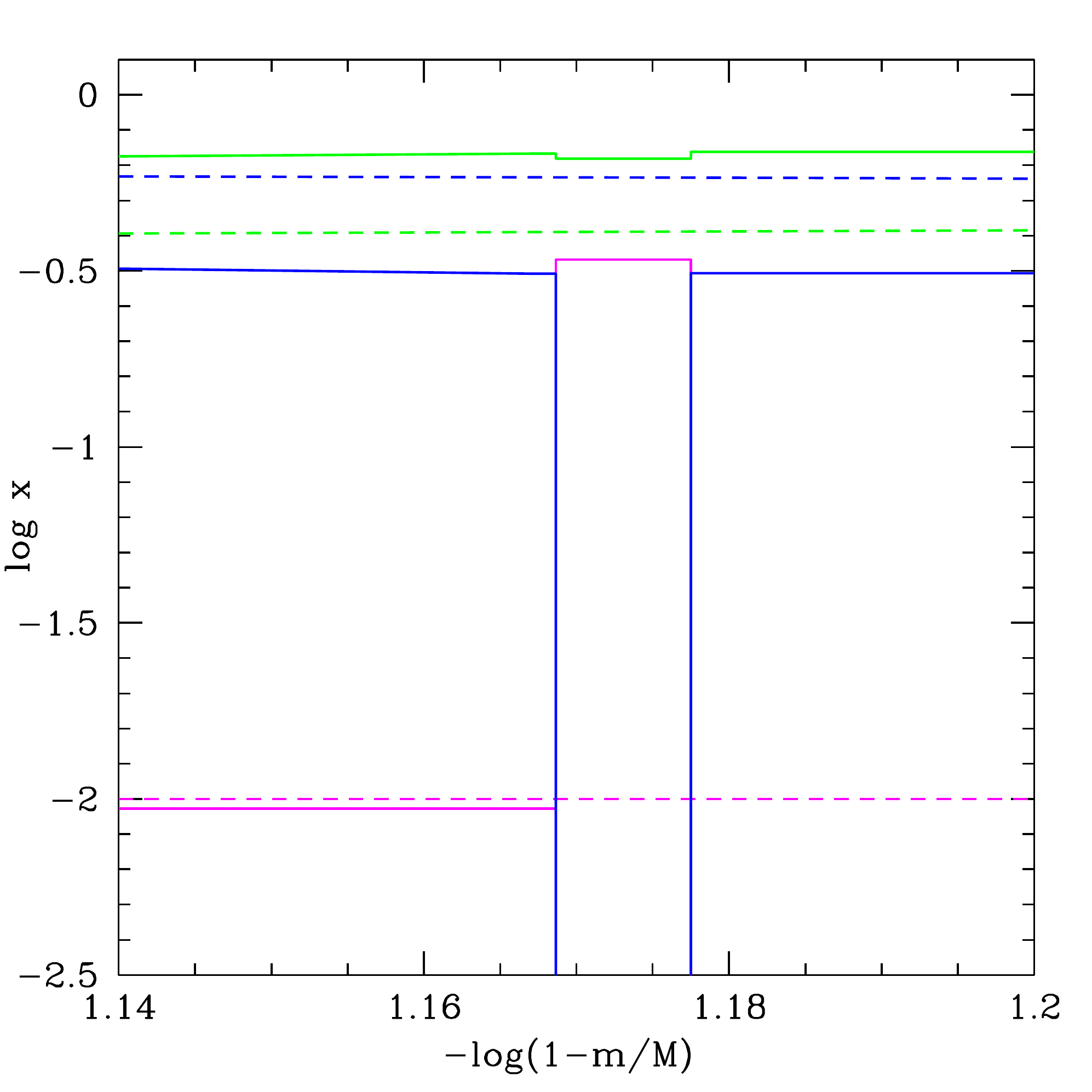}}
  \caption{The same meaning as Fig.~\ref{fig1}.} 
  \label{fig2}
\end{figure}

For the sake of simplicity, we are only going to consider the case of CO-white dwarfs. The inner part of the core has a composition dominated by the maximum extension of the central He-burning region where the abundance  of oxygen is maximum and almost constant. The size of this core increases with the mass of the white dwarf but, at the same time the abundance of oxygen decreases. Beyond this region the oxygen profile is built by the He-burning in a thick shell. Since gravitational contraction induces an increase of temperature and density, the abundance of oxygen decreases as a consequence of the fact that the ratio between the C-$\alpha$ and the 3$\alpha$ reaction rates  is lower for high temperatures, see Figure~\ref{fig1}, to the point that $^{12}$C becomes dominant \citep{dant89,sala97,sala00,sala10,fiel16}. This structure has been confirmed by the asteroseismological analysis of the ZZ Ceti star KIC08626021, although with noticeable surprises concerning the abundance and the size of the oxygen-core \citep{giam18}\footnote{see also the contributions of Charpinet and Giammichele in this volume.}.

At the end of the H-burning, almost all the initial content in $^{12}$C , $^{14}$N and $^{16}$O is converted into $^{14}$N which, in turn, is converted into $^{22}$Ne  during He-burning. The total abundance of this isotope is, thus, almost equal to the sum of the abundances of the other three ($x_{22}=x_{12}+x_{14}+x_{16}$) and becomes the most abundant of the minor species of the white dwarf (Figure\ref{fig1}. Besides this isotope and the initial content in metals, the parent star synthesised new isotopes during the different nuclear burning episodes as is the case of the s-elements which are synthesized during the AGB phase during double shell burning. These elements have a total abundance of $10^{-7}-10^{-6}$ in mass fraction in the outer 10\% of the white dwarf \citep[C. Abia, private communication and ][]{pier07}. To these impurities it is necessary to include the accretion of the circumstellar debris left by the progenitor.

It is usually assumed that once the white dwarf is born its chemical structure remains frozen. Assuming that the amount of accreted material is negligible, this is true for the total chemical content of the star since the nucleosynthesis activity is blocked by the electron degeneracy, but not for the distribution of the different elements along the star. As a consequence  of gravitational settling of neutron rich nuclei like  $^{18}$O, $^{22}$Ne, $^{56}$Fe, ... \citep{brav92,bild01,garc08}, and  sedimentation induced by a change of solubility upon crystallization \citep[][and cites there in]{iser98}, the chemical profile of white dwarfs experiences noticeable changes during the cooling process. 

For instance, consider of a white dwarf made of  $^{12}$C, $^{16}$O, and $^{22}$Ne only, as depicted in Figure~\ref{fig1}. The phase diagram of \citet{segr96} predicts that for small concentrations of neon, as far as the crystallization temperature satisfies the condition $T/T_{\rm C} \ge 1.12$, where $T_{\rm C}$ is the temperature of solidification of pure carbon, the liquid and the solid have the same abundance of neon and the phase diagram of the C/O mixture is not affected by the presence of neon. When this condition is not fulfilled, i.e, when the abundance of carbon is much more larger than that of oxygen, the presence of neon modifies  the phase diagram, which evolves into an azeotropic form,  and a process of distillation starts leading to the formation of shell made of carbon and neon only ($x_{\rm C} =0.78$, $x_{\rm O} =0.0$, and $x_{\rm Ne} =0.22$, by number). See figures 1 and 2, solid lines.

\section{The neon-E anomaly}
The neon-E anomaly was discovered by \citet{blac69} when they analyzed a sample from the Orgueil meteorite, a carbonaceous chondrite, using a step by step technique, and found that the neon released at high temperature (900 - 1100 $^o$C) was enriched in $^{22}$Ne as compared with the solar value. Subsequent studies lowered this value leading to the idea that Ne-E was made of pure $^{22}$Ne \citep{eber74,eber79,eber81,jung82}. These analysis also showed the existence of two types of Ne-E anomalies: the Ne-E(H), that is released at high temperatures (1200-1400 $^o$C) and is associated to high density minerals and the Ne-E(L) that is released at lower temperature (500 - 700 $^o$C) and is associated to low density minerals \citep{ande88}.  
It is specially significant that the carrier of the Ne-E(H) anomaly is silicon carbide \citep{bern87,tang88},  and that of Ne-E(L) is graphite \citep{amar90}. It is important to mention that all carriers containing noble gas are carbonaceous, and that graphite can only form in the absence of free oxygen.

The isotope $^{22}$Ne may have a \emph{thermonuclear} origin during the hydrostatic He-burning stage of stars, a \emph{radiogenic} origin as the decay product of the $^{22}$Na produced in the O/Ne zone of a supernova \citep{raus02,chie04} or in an ONe nova \citep{jose98,jose04}, or an \emph{spallative} origin by bombarding $^{25}$Mg with protons of moderate energy. The spallative origin has been rejected since it demands a very fine tuned pulse of protons \citep{audo76}. In the case of Ne-E(H) in presolar SiC grains, an extended idea is that it originated during the thermal pulses of AGB stars \citep{lewi90,lewi94, gall90, amar09} while the Ne-E(L) in graphite grains demands a more elaborated scenario. \citet{amar09} proposes that grains grow in the O-rich region where $^{22}$Na is produced and, thanks to the radioactive dissociation of the CO molecule, graphite can form according to \citet{clay99}. An additional possibility could be the ejection of the C/Ne layer formed as a consequence of the crystallization process if this layer survive to the collision. 

\section{White dwarf collisions}

The outcome of a collision between two stars essentially depends on three factors, the velocity of the collision, the structure of the stars, including the mass, and the impact parameter. As a general rule, frontal collisions  are the most efficient converting kinetic into thermal energy, and compact stars are less affected than extended stars \citep{shar02}.

There are few simulations of the collision of two white dwarfs \citep{benz89, rask09,rask10, ross09, lore09, azna13} and only recently, when it has been realized that the frequency of these events could be larger than expected thanks to the contribution of hierarchical triple systems, has started to increase.The outcome of these collisions can be a direct collision, with a unique mass transfer episode that disrupts the secondary in dynamical timescale, a lateral collision in which the secondary is totally disrupted after several orbits, and finally the formation of an eccentric binary system. Collisions with a small impact parameter can produce important amounts of $^{56}$Ni and can be assimilated to some Type Ia subtypes. Also, during such collisions, important amounts of intermediate mass elements (IMS) like $^{28}$Si are synthesized. In almost all cases, important amounts of matter, in the range of $0.01$ to $1$ M$_\odot$ are ejected. Therefore the question is to know if at least part of this layer can be ejected preserving its composition.

\section{Results and conclusions}

\begin{figure}[t]
  \centerline{\includegraphics[width=1.0\columnwidth]{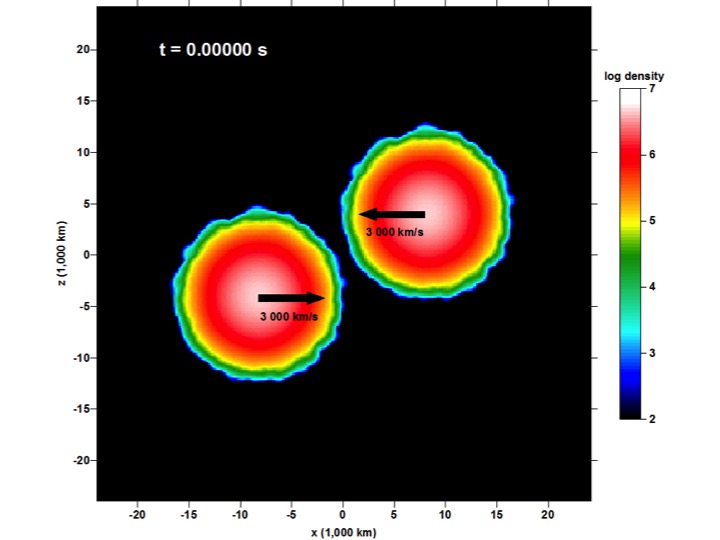}}
  \caption{ Density distribution of the two white dwarfs of 0.64 M$_\odot$ each one before the collision in the plane defined by the velocities of the centers of mass of each one.} 
  \label{fig3}
\end{figure}

\begin{figure}[t]
  \centerline{\includegraphics[width=1.0\columnwidth]{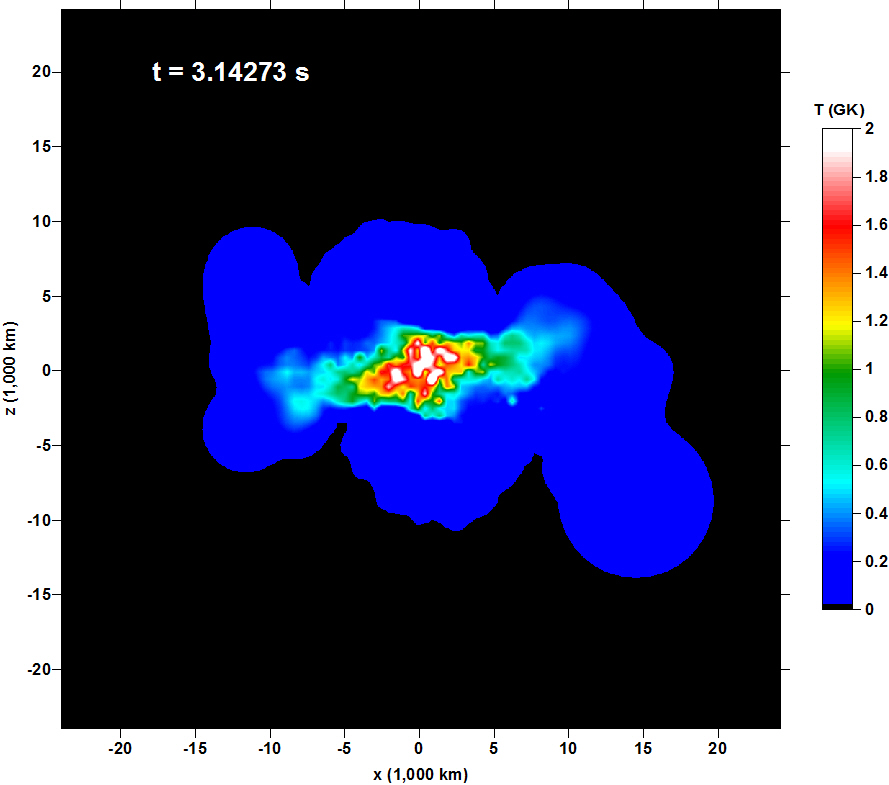}}
  \caption{Temperature distribution, in units of $10^9$K, in the same plane as in Fig.~\ref{fig3}, at the instant 3.14273 s, when the temperature at the point of impact is near its maximum value.} 
  \label{fig4}
\end{figure}

As a first step we have computed the collision of two C/O white dwarfs of 0.64 M$_\odot$, with a relative velocity of 6,000 km s$^{-1}$, and an impact parameter equal to the radius of the stars. The initial chemical composition is the same as in Figure~\ref{fig1}, which is similar to that obtained by \citet{cami16} for a 3 M$\odot$ main sequence star with solar metallicity. To simulate the event we have used the SPH models described in \citet{brav09a,brav09b} using $ 2.5\times 10^5$ particle for each star (see Figure~\ref{fig3})

Figure~\ref{fig4} displays the distribution of  temperatures at t=3.14273 s, when the temperature at the impact point is, approximately, maximum. As it can be seen important parts of the white dwarfs are preserved, but other are heated to temperatures high enough to synthesize  intermediate mass elements like $^{28}$Si or even $^{56}$Ni. The total amount of mass ejected during this first step of the collision is 0.064 M$_\odot$ containing $2.2\times 10^{-4}$ M$_\odot$ of C/Ne mixture as well as important amounts of other elements like, $1.1\times 10^{-2}$ M$_\odot$ of $^{28}$Si and $3.2\times 10 ^{-4}$ M$_\odot$ of $^{56}$Ni as well as $1.98\times 10^{-2}$ and $1.92\times 10^{-2}$ M$\odot$ of $^{12}$C and $^{16}$O respectively.

\begin{figure}[h]
  \centerline{\includegraphics[width=1.0\columnwidth]{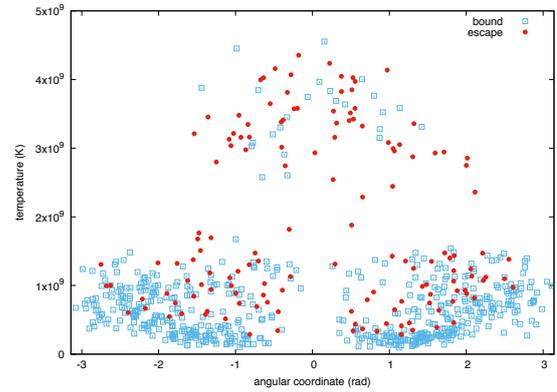}}
  \caption{Temperature reached by the different SPH particles in which the C/Ne layer was distributed versus the angular coordinate. Red dots represent the particles that escape and blue squares those that remain bounded. Below a temperature of $2\times10^9$~K no substantial carbon burning occur and the composition of the C/Ne layer is expected to experience only small modifications.} 
  \label{fig5}
\end{figure}

The average composition of the C/Ne mixture is, in mass fraction, x$_{12}=0.403$, x$_{22}=0.206$, x$_{20}=6.557\times 10^{-3}$. However the isotopic distribution can be different for the different particles since the maximum temperature reached by them is also different as it can be seen in Figure~\ref{fig5}. Therefore, the next step is to explore the large variety of parameters defining these systems to identify the possible anomalies  they can introduce and, in particular, to see if this scenario can totally or partially account those found in the Solar System.

\begin{figure}[t]
  \centerline{\includegraphics[width=1.0\columnwidth]{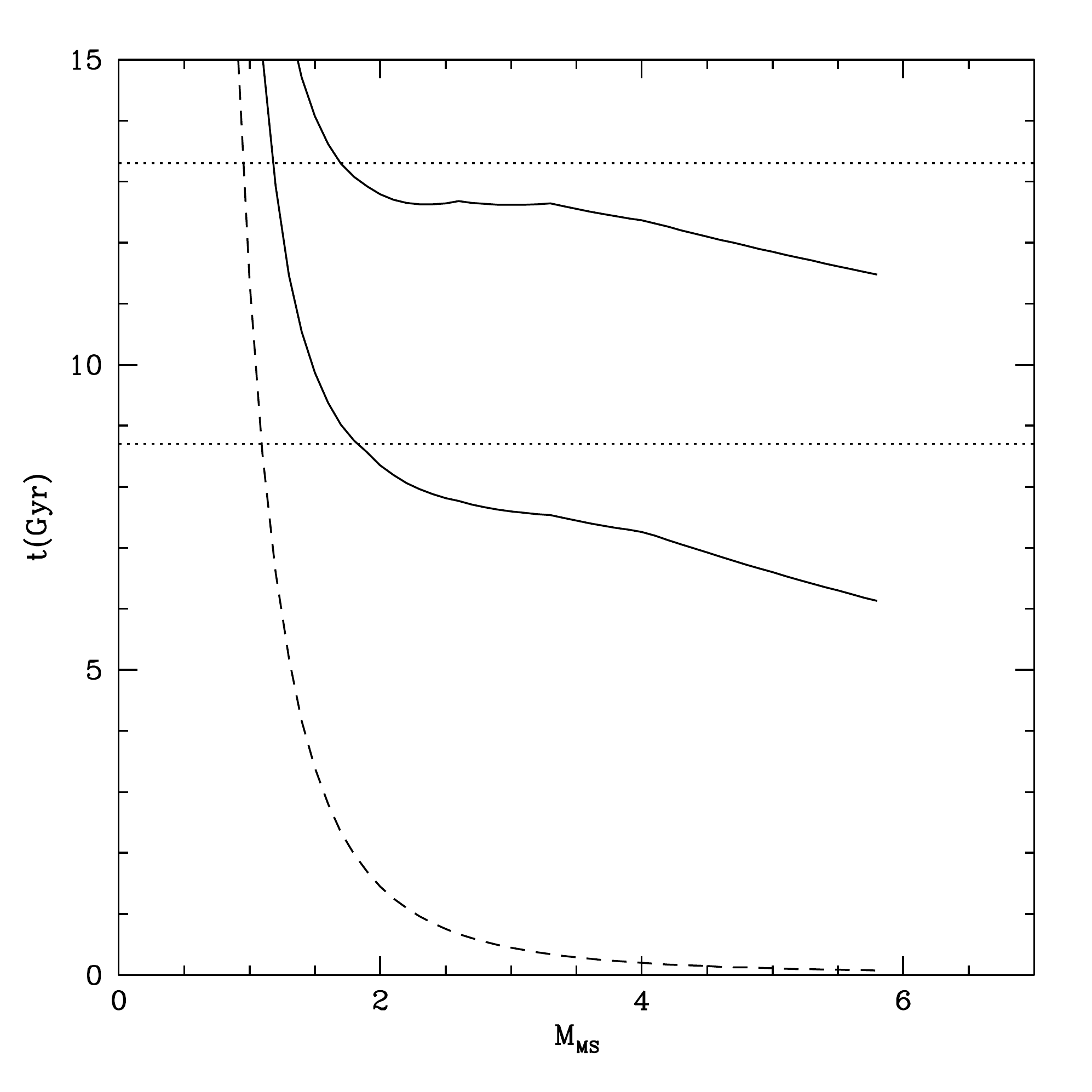}}
  \caption{Time to reach a luminosity $-\log(L/L_\odot \approx 4.4$ (cooling time plus lifetime of the progenitor) for different parent stars of DA and non-DA white dwarfs (up and low solid lines respectively). Dashed line, lifetime of the progenitor assuming solar metallicity. Upper dotted line, age of the Milky Way, lower dotted line, age of the Galaxy when the Solar System formed.} 
  \label{fig6}
\end{figure}

Concerning the Neon-E(L) anomaly, an important question is to know if there is time enough to guarantee the formation of the azeotropic layer. Figure~\ref{fig6} shows that non-DA white dwarfs can reach this stage of crystallization before the formation of the Solar System if their Main Sequence progenitors are massive enough.
It is interesting to remember that the detection in 2017 of \emph{Oumuamua},  a possible interstellar object, can remove this constrain and future missions to obtain samples of these objects could provide direct information about the behavior of matter in the deep interior of stars \citep{seli18}.
%------------------------------------------------

 Naturaly, before arriving to any conclusion, it is also necessary to examine what happens with the abundances of other isotopes found in the carriers, which is the mechanism of formation of dust in these collisions,if they can produce another anomalies ... and, finally they can provide important constraints to the rate at which these collisions can occur.

\section*{ Acnowledgements}
This work  has been supported  by MINECO grants  ESP2013-47637-P (JI) and ESP2015-63588-P (EB), by the  European Union FEDER funds, and by the grant 2014SGR1458 (JI)  of  the Generalitat de Catalunya.

%----------------------------------------------------------------------------------------
%	BIBLIOGRAPHY
%----------------------------------------------------------------------------------------

\bibliography{isern_euwd21}
% Here we have assumed that the bibliography file is named
% "papers.bib"

%----------------------------------------------------------------------------------------

\end{document}